\documentclass[pre,twocolumn,aps,eqsecnum]{revtex4}
\usepackage{epsfig}

\begin{document}

\title{Shear-Transformation-Zone Theory of Viscosity, Diffusion, and Stretched Exponential Relaxation in Amorphous Solids}

\author{J.S. Langer}
\affiliation {Department of Physics, University of California, Santa Barbara, CA  93106-9530}

\date{\today}

\begin{abstract}
The shear-transformation-zone (STZ) theory has been remarkably successful in accounting for broadly peaked, frequency-dependent, viscoelastic responses of amorphous systems near their glass temperatures $T_g$.  This success is based on the theory's first-principles prediction of a wide range of internal STZ transition rates.  Here, I show that the STZ rate-distribution causes the Newtonian viscosity to be strongly temperature dependent; and I propose that it is this temperature dependence, rather than any heterogeneity-induced enhancement of diffusion, that is responsible for Stokes-Einstein violations near $T_g$.  I also show that stretched-exponential relaxation of density fluctuations emerges naturally from the same distribution of STZ transition rates that predicts the viscoelastic behavior. To be consistent with observations of Fickian diffusion near $T_g$, however, an STZ-based diffusion theory somehow must include the cascades of correlated displacement events that are seen in low-temperature numerical simulations.
\end{abstract}
\maketitle

\section{Introduction}
\label{intro}

Among the deepest challenges in glass physics is understanding the temperature dependence of the Stokes-Einstein ratio in the neighborhood of the glass temperature $T_g$ \cite{EDIGER-06, EDIGER-09, BARTSCH-06}.  An apparently related phenomenon is the stretched-exponential relaxation of density fluctuations and other correlations.  Both phenomena have been cited as evidence for dynamic heterogeneities \cite{LEIDEN}, which supposedly provide rapid diffusion paths relevant to the first case, and a variety of localized, relaxation environments in the second \cite{EDIGER-00}.  Here, I offer a counterproposal -- that these phenomena can be explained more systematically by the multi-species shear-transformation-zone (STZ) theory that Bouchbinder and I  \cite{BL-PRL-11,BL-PRE-11} invoked to understand the broad range of time scales observed in viscoelastic experiments. A short account of this proposal has appeared in \cite{JSL-arXiv-11}.

The present version of the STZ theory is an extension of the flow-defect theories of Turnbull, Cohen, Spaepen, Argon and others \cite{TURNBULL-COHEN-70,SPAEPEN-77,ARGON-79,SPAEPEN-81,ARGON-83} in which localized clusters of molecules undergo irreversible rearrangements in response to applied shear stresses. In more recent publications \cite{FL-98,JSL-08,FL-11}, my coworkers and I have argued that these flow defects, i.e. the STZ's, must be dynamic entities.  They are thermally activated, structural fluctuations that appear and disappear on time scales of the order of the $\alpha$ relaxation time $\tau_{\alpha}$.  In the presence of a shear stress, they are the places where irreversible molecular  rearrangements occur.  

The STZ's also may be associated with self diffusion of a tagged molecule.  At most times, such a molecule remains in its cage, surrounded by its neighbors. It participates in a rearrangement involving these neighbors only when a thermally activated structural fluctuation -- a shear transformation -- occurs at its location.  When that happens, the tagged molecule jumps some distance, perhaps an intermolecular spacing $a$, in some direction, while one or more of its neighbors jumps in the other direction. A sequence of such jumps becomes a diffusive trajectory. 

In almost all of the earlier STZ literature, for example in \cite{FL-11}, the main focus has been on nonlinear phenomena.  We have computed the quasistationary responses of amorphous materials to strong, slow, driving forces, and have argued that the configurational states of such systems must necessarily be described by the effective temperatures of their structural degrees of freedom \cite{JSL-08,FL-11,BLII-09,BLIII-09}.  Here, in contrast, I focus on well equilibrated glassy systems, near their glass temperatures, subject only to infinitesimally weak perturbations.  Newtonian linear viscosity, self diffusion, and time dependent relaxation of small density fluctuations as seen in scattering experiments, all fall into this category. In this linear limit, the effective temperature is the same as the ambient temperature, and the noisy fluctuations that activate the creation and annihilation of STZ's are purely thermal.  Therefore, this paper and its immediate predecessors \cite{BL-PRL-11,BL-PRE-11} probe different aspects of glass physics than  usually have been discussed in connection with STZ theory.

The most important way in which these papers differ from the earlier ones is in their introduction of multiple species of STZ's.  In  \cite{BL-PRL-11,BL-PRE-11}, Bouchbinder and I argued that the STZ's in well equilibrated, low-temperature systems predictably occur with a broad range of internal transition rates, and that the resulting multi-species  theory accurately accounts for experimentally observed, frequency-dependent, viscoelastic response functions \cite{GAUTHIER-04}.  In Sec.\ref{STZ} of this paper, I briefly summarize the linearized equations of motion for this class of STZ theories. Then, in Sec.\ref{stokes-einstein}, I  show that this multi-species theory predicts a strongly temperature dependent viscosity that, even in the absence of enhanced diffusion, is consistent with the observed Stokes-Einstein violations.  The extension of this theory to self-diffusion is discussed in Sec.\ref{self-diffusion}. There, I show that the presence of slowly responding STZ's in the multi-species theory requires the existence of something like cascades of correlated hopping events in order that the theory be consistent with observations of normal Fickian diffusion near the glass temperature.  I also show that the distribution of STZ transition rates predicts stretched-exponential decay of density fluctuations in apparent agreement with observations. I conclude in Sec.\ref{conclusions} with remarks about ways in which these theoretical proposals need to be tested experimentally.

\section{Linear STZ Dynamics}
\label{STZ}

The STZ theory for amorphous systems in thermodynamic equilibrium near their glass temperatures is concisely summarized  by the master equation for the number densities $n_{\pm}(\nu)$ of STZ's oriented parallel or antiparallel to the shear  stress $s$:
\begin{eqnarray}
\label{ndot}
\nonumber
\dot n_{\pm}(\nu) &=& R(\pm s)\,n_{\mp}(\nu) - R(\mp s)\,n_{\pm}(\nu)\cr\\  &+& \rho(\theta)\,\bigl[n_{eq}(\theta) - n_{\pm}(\nu)\bigr].
\end{eqnarray}
Here, times are measured in microscopic units, perhaps molecular vibration periods of the order of picoseconds. The $R(\pm s)$ are the rates of forward and backward STZ transitions.  The symbol $\nu$ denotes the internal transition rate, defined for this purpose to be $\nu \equiv 2\,R(0)$.  Writing Eq.(\ref{ndot}) with this explicit $\nu$ dependence implies that a range of different species of STZ's, with different values of $\nu$, will be relevant to the phenomena of interest.  In the following discussion, I denote the probability distribution for the $\nu$'s by $\tilde p(\nu)$. 

The second pair of terms on the right-hand side of Eq.(\ref{ndot}) are the rates of STZ creation and annihilation. These are thermally activated processes, expressed in the form of a detailed-balance relation in which the equilibrium  STZ density is 
\begin{equation}
\label{neq}
n_{eq}(\theta) = {1\over v_0}\,e^{-e_Z/\theta}.
\end{equation}
where $\theta = k_BT$ is the temperature in energy units.  $e_Z$ is the STZ formation energy which, to a first approximation, is the energy required to create the amorphous analog of a vacancy-interstitial pair.  $v_0$ is the average volume per molecule, proportional to the cube of the average molecular spacing $a$.

The attempt frequency $\rho(\theta)$ is best understood as a dimensionless, thermal noise strength. It is a super-Arrhenius function whose strong $\theta$ dependence governs the equilibrium glass transition.  Because time is measured  in molecular units, $\rho(\theta)$ must be of the order of unity at large $\theta$, where the attempt frequency is the molecular vibration frequency. As $\theta$ decreases through the glass temperature $\theta_g = k_B T_g$, the rate of thermally activated structural rearrangements slows dramatically, and $\rho(\theta)$ falls rapidly toward zero. The structural relaxation rate, $\rho(\theta)\,\exp\,(-e_Z/\theta)$, is conventionally identified as the $\alpha$ relaxation rate $\tau_{\alpha}^{-1}$. For present purposes, I take $\rho(\theta)$ to be an observable but not necessarily  predictable quantity. See \cite{Langer-PRE-06,Langer-PRL-06} for my best attempt to date to compute it from first principles.

All of the phenomena to be considered here are technically ``slow'' in the sense that their characteristic time scales are much longer than the times on which the $n_{\pm}(\nu)$ in Eq.(\ref{ndot}) relax to quasistationary values.  Note that this was not true for the oscillatory viscoelasticity discussed in \cite{BL-PRL-11,BL-PRE-11}, where the broad range of driving frequencies included those that are fast as well as slow on intrinsic STZ time scales. Here, we are interested only in linear steady-state viscous flow, or in diffusional relaxation over times of the order of or longer than $\tau_{\alpha}$; therefore, all of the information that we need can be obtained by setting $\dot n_{\pm}(\nu) = 0$ on the left-hand side of Eq.(\ref{ndot}) and solving for the $n_{\pm}(\nu)$.  

In this quasistationary approximation, to first order in the stress $s$, the steady-state plastic strain rate $\dot\gamma^{pl}$ has the form
\begin{eqnarray}
\label{gammadot}
\nonumber
\dot\gamma^{pl}&=& \epsilon_0\,v_0\,\int_0^{\infty} d\nu\,\tilde p(\nu)\,\left[R(s)\,n_{-}(\nu) - R(- s)\,n_{+}(\nu)\right]\cr \\&=& \epsilon_0\,\int_0^{\infty} d\nu\,\tilde p(\nu)\,{1\over \tau(\nu)}\,{\cal T}(s),
\end{eqnarray}
where
\begin{equation}
\label{tau}
{1\over \tau(\nu)}= {1\over \tau_{\alpha}(\theta)}\,\left({\nu\over \nu + \rho}\right), ~~~ {1\over \tau_{\alpha}(\theta)}\equiv \rho(\theta)\,e^{-\,e_Z/\theta},
\end{equation}
and, to first order in $s$, 
\begin{equation}
\label{calT}
\nu \equiv 2\,R(0),~~~~{\cal T}(s)\equiv {R(s) - R(-s)\over R(s) + R(-s)} \approx {v_0\,s\over \theta}.
\end{equation}
Equation(\ref{gammadot}) comes directly from Eq.(\ref{ndot}) with no additional assumptions.  Its structure implies that $\tau(\nu)^{-1}$ is the event rate for STZ's of species $\nu$.  As explained in Appendix \ref{appendix}, $\tau(\nu)^{-1}$ is the frequency at which an STZ is created in one of its internal states and annihilated in the other state, having made an arbitrary (odd) number of transitions between those states during its lifetime.  

There are several aspects of these equations that need emphasis.  First, they assume that $\epsilon_0\,v_0$, the volume of the plastic core of a shear transformation, is independent of $\nu$. Second, the definition of $\tau_{\alpha}$ introduced earlier and again in Eq.(\ref{tau}), is not tied directly to the temperature dependence of the viscosity $\eta$.  That is,  $\eta$ is not always simply proportional to $\tau_{\alpha}(\theta)$. 

Third, ${\cal T}(s)$ is the stress-induced bias between forward and backward transitions.  The final expression in Eq.(\ref{calT}) is the classic Einstein formula, deduced from the assumption that the underlying transition is a stress-enhanced, thermally activated process.  In \cite{BLIII-09}, this result was shown to follow from the second law of thermodynamics, with $\theta$ more generally being the thermodynamically defined effective temperature $\chi$.  A simple application of the preceding analysis to rapidly sheared systems, where the transverse diffusion constant is $a^2/\tau_{\alpha}$, confirms that $\chi$ is the same as the effective temperature determined by a fluctuation-dissipation relation, at least within the STZ theory. (See \cite{CUGLIANDOLO-11} and references cited there.)  Throughout the present paper, thermodynamic equilibrium implies that $\chi = \theta$.

In \cite{BL-PRL-11,BL-PRE-11}, Bouchbinder and I derived a formula for the distribution $\tilde p(\nu)$ starting with the assumption that the zero-stress transition rate has the form
\begin{equation}
\label{nudef0}
\nu = 2\,R(0) = \rho_0\,e^{- \Delta/\theta},
\end{equation}
where $\Delta$ is an activation energy. $\rho_0$, in analogy to $\rho$ in Eq.(\ref{ndot}), is an attempt frequency that, in principle, can depend on both $\Delta$ and $\theta$.  Continuing this analogy, we argued that $\rho_0$ must be unity for small $\Delta$ and large $\theta$ (fast transitions), but becomes small in the opposite limit.  Because $\Delta$ is measured {\it downward} from some reference energy, its probability distribution has the form
\begin{equation}
\label{pDelta}
p(\Delta) \propto e^{+ \zeta\,\Delta/\theta},
\end{equation}
where $\zeta$ is a positive, $\theta$-independent constant that is less than unity.  Then, for small $\Delta$ (large $\nu$), 
\begin{equation}
\tilde p(\nu) = p(\Delta)\,\left|{d\Delta\over d\nu}\right| \propto {1\over \nu^{1+\zeta}}.
\end{equation}
This distribution diverges at small $\nu$ and must be cut off there.  To choose that cutoff, say at $\nu = \nu^*$, we argued that the internal transition rates cannot be slower than the rate of spontaneous structural rearrangements. Therefore, we chose $\nu^* = \tau_{\alpha}^{-1}$, which, according to Eq.(\ref{tau}), is several orders of magnitude  smaller than $\rho$ for systems near their glass temperatures. The corresponding distribution over barrier heights, $p(\Delta)$, as given in Eq.(\ref{pDelta}) with a sharp cutoff at large $\Delta$, is roughly consistent with that found experimentally by Argon and Kuo \cite{ARGON-KUO-80}.  

Using these constraints, Bouchbinder and I wrote a three-parameter $\nu$ distribution in the form
\begin{equation}
\label{pnu}
\tilde p(\nu) = {\tilde A\over \nu\,[(\nu/\nu^*)^{\zeta} + (\nu^*/\nu)^{\zeta_1}]},
\end{equation}
where $\tilde A$ is a normalization constant.  We used $\zeta = 0.4$ in accord with  viscoelastic data for various materials near their glass temperatures \cite{GAUTHIER-04}. The exponent $\zeta_1$, which determines the sharpness of the cutoff at $\nu^*$, is not sharply determined by experiment. For most purposes, we have used $\zeta_1 = 1$. There was no need to specify the quantity $\rho_0$ in Eq.(\ref{nudef0}); its principal role is to provide a physical mechanism for producing very small values of $\nu$ without requiring unphysically large values of $\Delta$. These first-principles estimates of $\tilde p(\nu)$ and $\nu^*$, with no additional fitting parameters, are accurately confirmed by the experimental data.  

\begin{figure}[here]
\centering \epsfig{width=.45\textwidth,file=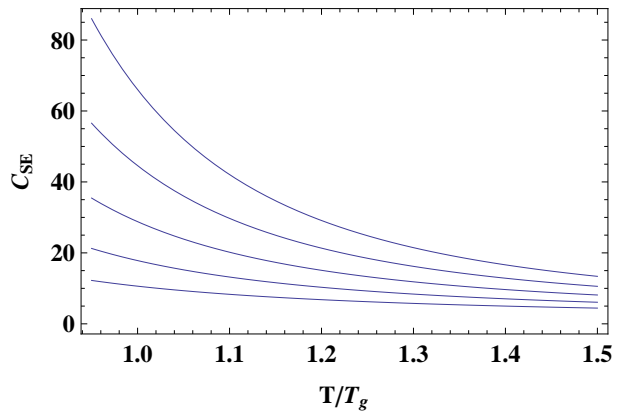} \caption{Stokes-Einstein ratio $C_{SE}$ for $\zeta$ = 0.4, 0.5, 0.6, 0.7, and 0.8 from bottom to top.} \label{CSE(T)}
\end{figure}

\section{Viscous Flow and the Stokes-Einstein Ratio}
\label{stokes-einstein}

A central feature of the present analysis is that, as implied by Eq.(\ref{tau}) and discussed in the Appendix, there are two different limiting populations of STZ's at low temperatures.  There are ``fast'' STZ's, with $\nu \gg \rho$, for which  $\tau(\nu) \approx \tau_{\alpha}$.  These STZ's may make multiple, back and forth transitions between their two states during their lifetimes.  There is also a substantial population of ``slow'' STZ's with $\nu \ll \rho$ and $\tau(\nu) \approx (\rho/\nu)\,\tau_{\alpha} \gg \tau_{\alpha}$, which make at most one transition before disappearing.  Both populations are important for determining the inverse viscosity:
\begin{equation}
\label{eta}
{1\over \eta(\theta)} = {\dot\gamma^{pl}\over s} = {\epsilon_0\,v_0\over \theta\,\tau_{\alpha}(\theta)} \int_0^{\infty} d\nu\,\tilde p(\nu)\,\left({\nu\over \nu + \rho}\right).
\end{equation}

Equation (\ref{eta}) makes it clear that the distribution $\tilde p(\nu)$ plays a crucial role in determining the viscosity.  At high temperatures, $\nu^*$ is large, of the order of unity, and $\tilde p(\nu)$ must be peaked at a comparably large value of $\nu$.  As a result, Eq.(\ref{eta}) becomes a conventional formula in which $\eta \propto \theta\,\tau_{\alpha}(\theta)$.  On the other hand, at temperatures near or below $\theta_g$, the integration in Eq.(\ref{eta}) is dominated by the slow STZ's with $\nu \approx \nu^*$, and the conventional formula for $\eta$ is enhanced by a factor that can be shown to be roughly proportional to $\exp\,(\zeta/\theta)$.

Suppose, for the moment, that diffusion in this system is entirely normal, with a diffusion constant given simply by ${\cal D} = a^2/\tau_{\alpha}$.  Then Eq.(\ref{eta}) predicts that the Stokes-Einstein ratio is
\begin{equation}
\label{SE}
{a_0\,{\cal D}\,\eta\over \theta} = \left[\int_0^{\infty} d\nu\,\tilde p(\nu)\,\left({\nu\over \nu + \rho}\right)\right]^{-1} \equiv C_{SE}(\theta),
\end{equation}
where $a_0 \equiv \epsilon_0\,v_0/a^2$.  Since  $\tau_{\alpha}$ has cancelled out of Eq.(\ref{SE}), and $\tilde p(\nu)$ is a function only of $\nu/\nu^*$ (apart from a cutoff at large $\nu$ when necessary), the only temperature dependence of $C_{SE}(\theta)$ occurs via the ratio $\nu^*/\rho \sim  \exp(-e_Z/\theta)$. In \cite{BL-PRL-11,BL-PRE-11}, we estimated that $\exp(-e_Z/\theta_g) \sim 10^{-3}$.  That estimate has been used in evaluating Eq.(\ref{SE}) to plot the graphs of $C_{SE}$ as functions of $T/T_g$ shown in Fig. \ref{CSE(T)}.  The balance between the contributions of slow and fast STZ's in Eq.(\ref{SE}) is determined largely by the exponent $\zeta$ in Eq.(\ref{pnu}), thus $C_{SE}$ is shown in the figure for five different values of $\zeta$.  This theory does not (yet) have enough microscopic physical content to make an accurate  connection between low and high-temperature behaviors, or even to make material-specific predictions of the shape of $\tilde p(\nu)$ or the length scale $a_0$. If, as argued in the preceding paragraph, $\tilde p(\nu)$ becomes peaked at large values of $\nu$ for high temperatures, then $C_{SE}\cong 1$ in that limit.  Thus, the values of  $C_{SE}(\theta_g)$ in Fig. \ref{CSE(T)} imply Stokes-Einstein violations of roughly the magnitude seen experimentally, arising entirely from the anomalously large viscosity predicted by Eq.(\ref{eta}).  

\section{Self-Diffusion}
\label{self-diffusion}

Turn now to the statistical problem of self diffusion.  Here, instead of averaging over probabilities of single events as above, we must consider sequences of multiple hopping events. The statistical weight assigned to any such sequence depends on how far the molecule moves and how long it takes to get there.  Technically, the molecule executes  a ``continuous-time random walk'' (CTRW) \cite{MONTROLL-SHLESINGER-84,BOUCHAUD-GEORGES-90}. Note that the model of STZ-enabled diffusion steps contains no information about dynamics on sub-molecular space or time scales.  In effect, it is coarse grained on the scale of the molecular spacing, and it ``knows'' nothing about fast vibrational motions other than that they carry the thermal noise that activates barrier-crossing transitions. Therefore, this model cannot describe $\beta$ relaxation or any details of how a molecule moves within the cage formed by its neighbors, or of how it escapes from that cage. 

The basic ingredient of the CTRW analysis is the one-step probability that, after waiting a time $t$, an STZ fluctuates into existence at the position of the tagged molecule, and the molecule jumps  a distance $z$.  (There is no loss of generality in projecting the diffusion mechanism onto one dimension.)  For an STZ with an internal rate $\nu$, this normalized one-step probability distribution is 
\begin{equation}
\psi(t,z,\nu) = f(z,\nu)\,{1\over\tau(\nu)}\,e^{-\,t/\tau(\nu)}, 
\end{equation}
where $f(z,\nu)$ is the probability of a jump of length $z$. As indicated, $f(z,\nu)$ may depend on $\nu$. The crucial assumption here is that the time $\tau(\nu)$ appearing in this formula is the same as the quantity defined in Eq.(\ref{tau}).  In both cases, $\tau(\nu)^{-1}$ is the average rate at which STZ's appear, make one net internal transformation (perhaps after multiple back-and-forth transformations), and then disappear.  This elementary fluctuation mechanism should be common to both plastic deformation and self diffusion.

The probability of any random walk consisting of a sequence of such steps is a product of space-time convolutions of factors $\psi(t,z,\nu)$. If the STZ events are uncorrelated -- a major assumption -- then each factor can be averaged independently over $\nu$.  The walk ends after a final time interval $t$ within which the molecule does not make a further jump; thus the complete path probability contains one factor 
\begin{equation}
\phi(t,\nu)={1\over\tau(\nu)}\,\int_t^{\infty}dt' \,e^{-\,t'/\tau(\nu)}= e^{-\,t/\tau(\nu)},
\end{equation} 
again included in the convolution integrals and averaged over $\nu$.

These space and time convolutions are converted into products by computing Fourier and Laplace transforms, in terms of which the multi-step walk probabilities can be summed to all orders. The self-intermediate scattering function $\hat F(k,t)$, i.e. the $k$'th Fourier component of the diffusion profile, is given by an inverse Laplace transform:
\begin{equation}
\label{Fhat-w}
\hat F(k,t) = \int_{\delta-\,i\,\infty}^{\delta +\,i\,\infty} {dw\over 2\,\pi\,i}\, {e^{w\,t/\tau_{\alpha}}\,\tilde K(w)\over 1 - \tilde J(k,w)},~~~\delta > 0,
\end{equation}
where
\begin{equation}
\label{Kdef}
\tilde K(w)= \int_0^{\infty} d\nu\,{\tilde p(\nu)\over w + \lambda(\nu)},~~\lambda(\nu) = {\tau_{\alpha}\over \tau(\nu)} = {\nu\over \nu + \rho}      
\end{equation} 
is the Laplace transform of the averaged $\phi(t,\nu)$, and 
\begin{equation}
\label{Jdef}
\tilde J(k,w) = \int_0^{\infty} d\nu\,\tilde p(\nu)\,{\lambda(\nu)\,\hat f(k,\nu)\over w + \lambda(\nu)}
\end{equation}
is the analogous transformation of $\psi(t,z,\nu)$. 

A crucial ingredient in these formulas is the jump-length distribution $\hat f(k,\nu)$. If, according to the discussion following Eq.(\ref{calT}), each STZ event moves the tagged molecule a mean-square distance $a^2$, then it would be natural to choose a Gaussian distribution $\hat f(k,\nu)= \exp(- k^2a^2/2)$ independent of $\nu$.  With this choice, however, the slow STZ's play the role of deep traps \cite{BOUCHAUD-GEORGES-90}, effectively shutting down long-time diffusion. To see this, compute the mean-square displacement $\overline{z^2(t)}$, which, for a $\nu$-independent $\hat f(k)$, becomes
\begin{equation}
\label{zsquared}
\overline{z^2(t)} = -\,\left[{\partial^2 \hat F(k,t)\over \partial k^2}\right]_{k=0}= a^2\,\int {dw\over 2\,\pi\,i}\, {e^{w\,t/\tau_{\alpha}}\over w^2\,\tilde K(w)}.
\end{equation} 
If $\tilde p(\nu)$ cuts off sharply at small $\nu$, then $K(w)$ is analytic at $w=0$, and the long-time behavior is obtained by integrating around the pole there.  The result is that $\overline{z^2(t)} \approx {\cal D}_0\,t$, with  ${\cal D}_0 = (a^2/\tau_{\alpha})\,\tilde K(0)^{-1}$.  Therefore, ordinary diffusion is suppressed by a factor $\tilde K(0)^{-1}\sim \nu^*/\rho \ll 1$. For $\zeta_1 \le 1$ in Eq.(\ref{pnu}), $\tilde K(0)$ diverges, and ${\cal D}_0$ vanishes.  This result is manifestly inconsistent with the experimental data reported in \cite{EDIGER-06, EDIGER-09, BARTSCH-06}, where Fickian diffusion was observed near the glass temperature.  

\begin{figure}[here]
\centering \epsfig{width=.45\textwidth,file=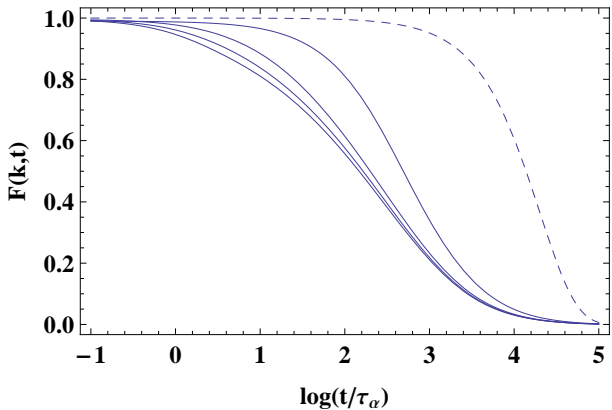} \caption{Computed self-intermediate scattering functions for $ka$ = 2.0, 1.0, 0.5, and 0.1, from left to right. The corresponding stretched-exponential indices are $b \cong$  0.43, 0.45, 0.58, and 0.78. The dashed curve is the Fickian limit for $ka = 0.01$. Only $\alpha$ relaxation appears here.  Space and time scales relevant to $\beta$ relaxation have been averaged out.} \label{SER-1}
\end{figure}

Given the $\tilde p(\nu)$ determined by the frequency-dependent viscoelasticity  \cite{BL-PRL-11,BL-PRE-11}, this discrepancy between diffusion theory and experiment appears to falsify the postulate of uncorrelated short jumps within the context of the present theory. It seems, instead, to indicate that the cascades of correlated STZ ``flips'' seen in low-temperature numerical simulations (see \cite{CAROLI-LEMAITRE-11} and earlier papers cited there) are relevant to laboratory systems at temperatures near $T_g$.  A systematic generalization of the CTRW analysis to include strongly correlated sequences of jumps is beyond the scope of this investigation; but it is plausible that such sequences might appear approximately in the form of anomalously long jumps initiated by individual slow events. Unlike the fast STZ's, the slow ones are stable only in near-equilibrium glass formers at low temperatures, and they only occasionally undergo shear transformations during their lifetimes. Their small values of $\nu$ imply that their internal energy barriers are high; therefore, when they do make  transitions, they release substantial amounts of energy that could trigger subsequent events.  Suppose, as a first guess, that a slow event sends a tagged molecule on a random walk that lasts for a time $\tau(\nu)$, during which the jump rate is $\tau_{\alpha}^{-1}$. The mean-square displacement during this walk would be  $a^2\,\tau(\nu)/\tau_{\alpha}$, which is equal to $a^2$ for fast STZ's, but becomes substantially larger for slow ones.  

The corresponding jump-length distribution is
\begin{equation}
\label{fhat}
\hat f(k,\nu) = e^{- k^2a^2/2\,\lambda(\nu)} = e^{- k^2a^2 \,\tau(\nu)/2\,\tau_{\alpha}}.
\end{equation}
This choice of $\hat f(k,\nu)$, if correct, would solve the problem of overly slow diffusion.  When Eq.(\ref{fhat}) is used in Eq.(\ref{Fhat-w}), the quantity $\tilde K(w)$ in the denominator of Eq.(\ref{zsquared}) cancels out, and  $\overline{z^2(t)} = a^2\,t/\tau_{\alpha}$ at all times. This means that diffusion in this approximation is normal in the sense that ${\cal D} = a^2/\tau_{\alpha}$, and that the assumption leading to the expression for the Stokes-Einstein ratio in Eq.(\ref{SE}) would be correct.  

Finally, consider the self-intermediate scattering function $\hat F(k,t)$ given by  Eqs.(\ref{Fhat-w}-\ref{Jdef}).  To evaluate $\hat F(k,t)$ numerically, close the contour of integration in Eq.(\ref{Fhat-w}) around the branch cut on the negative real $w$ axis.  Some results are shown in Fig.\ref{SER-1}.  The parameters correspond roughly to a metallic glass near $T_g$: $\rho/\nu^* = 10^3$, $\zeta= 0.4$, and $\zeta_1 = 1$.  The wavenumbers are $k\,a = 2.0,\,1.0,\,0.5,\,{\rm and}\, 0.1$, from left to right.  All four curves exhibit stretched-exponential relaxation of the form $\exp\,[- \,c\,(t/\tau_{\alpha})^b]$, with $b \cong 0.43,\,0.45,\,0.58,\,{\rm and}\,0.78$, in the same order of decreasing $k$. 

Several features of these results are notable. First, there is a Fickian limit at small $k$, where $b$ approaches unity, and $\hat F(k,t) \sim \exp\,(-\,{\cal D}\,k^2\,t/2)$ as shown by the dashed line in Fig.\ref{SER-1} for $ka=0.01$.  Second, there is a non-Fickian limit at large $k$, where the curves begin to lie on top of each other, and $b$ approaches $0.4$.  This limiting behavior occurs because any realistic choice of the factor $\hat f(k,\nu)$ vanishes at large $k$, leaving the initial, $k$-independent function $\tilde K(w)$ in the numerator of Eq.(\ref{Fhat-w}) as the only contribution to the relaxation function.  Note that 
\begin{equation}
\label{Kdef2}
K(t) \equiv \int {dw\over 2\,\pi\,i}\, e^{w\,t/\tau_{\alpha}}\,\tilde K(w)= \int_0^{\infty} d\nu\,\tilde p(\nu)\,e^{-\,t/\tau(\nu)}
\end{equation}
has the form of a local relaxation function averaged over a distribution of environments with different relaxation times $\tau(\nu)$. The transition between large-$k$ and small-$k$ behavior is temperature dependent {\it via} the quantity $\lambda(w)$ in Eq.(\ref{Fhat-w}), which produces a $\theta$ dependence here in much the way it did for the viscosity in Eq.(\ref{eta}).  At fixed $k$, according to these equations, the system changes from stretched-exponential to Fickian as the temperature increases.  It is not a coincidence that, in the large-$k$ limit given in Eq.(\ref{Kdef2}) and at low temperatures, $b \cong \zeta$; but $b = \zeta$ is not even an exact result for all choices of $\zeta$ and $\zeta_1$, nor is it exact in the limit of long times.

\section{Concluding Remarks}
\label{conclusions}

In summary, the multi-species STZ theory, with the transition-rate distribution $\tilde p(\nu)$ derived directly from statistical principles, and with essentially no arbitrary parameters, has been convincingly successful in predicting frequency-dependent viscoelastic responses of glassy materials.  As argued here in Sec.\ref{stokes-einstein}, this theory predicts that the steady-state, linear viscosity is strongly temperature dependent in the neighborhood of the glass temperature.  That prediction, in turn, suggests that  observed anomalies in the temperature dependence of the Stokes-Einstein ratio are due to large viscosities, and not necessarily to the emergence of rapid diffusion paths at low temperatures. The multi-species theory also predicts stretched-exponential decay of density fluctuations, and further predicts temperature and wave-number dependences of the stretched-exponential function that appear to agree with observations.  

The outstanding uncertainty in this theoretical picture is that, in its simplest version, it predicts that the slow STZ's act like traps and shut down Fickian diffusion at low temperatures, in contradiction to experimental measurements  \cite{EDIGER-06, EDIGER-09, BARTSCH-06}. To resolve this discrepancy, I have suggested in Sec.\ref{self-diffusion} that the correlated cascades of STZ-like events observed in low-temperature molecular-dynamics simulations \cite{CAROLI-LEMAITRE-11} need somehow to be incorporated into the diffusion analysis. This is done here, in an admittedly {\it ad hoc} way, by choosing a $\nu$-dependent jump size distribution $\hat f(k,\nu)$ in Eq.(\ref{fhat}). It is not clear whether Eq.(\ref{fhat}) is a realistic approximation, or whether a more accurate description of correlated diffusion steps might modify the predicted Stokes-Einstein ratio.

The analysis presented here calls for a unified investigation of Stokes-Einstein violations, stretched-exponential relaxation, and frequency dependent viscoelastic response functions, all in comparable glass-forming materials near their glass temperatures.  Checking the consistency of the various formulas shown here would test the theory as a whole and its interpretation of diverse dynamic phenomena.  For example, it ought to be possible to measure viscoelastic responses (or internal-friction functions) using the same materials, under the same conditions, that are used for measuring $\hat F(k,t)$ in the large-$k$ limit shown in Eq.(\ref{Kdef2}).  Then, it should be possible either to deduce consistent approximations for $\tilde p(\nu)$, or else to falsify the predicted consistency.  Similarly, it would be useful to make the latter kinds of measurements for the same systems in which Fickian diffusion is observed, and thus to test the approximation for $\hat f(k,\nu)$ in Eq.(\ref{fhat}).  Detailed studies of the $k$ and $\theta$ dependences of $\hat F(k,t)$, again on the same systems for which other measurements are made, might further resolve the outstanding issues. 

\appendix
\section{STZ Event Rate}
\label{appendix}

To interpret the event rate $\tau(\nu)^{-1}$ defined in Eq.(\ref{tau}), it is useful to perform a continuous-time, random-walk analysis similar to that used for diffusion in Sec.\ref{self-diffusion}, but now restricted to transitions back and forth between the internal $\pm$ states of a single STZ. 

Consider an elementary process in which an STZ is created in a $\mp$ state and then, after a sequence of back and forth transitions, is annihilated in a $\pm$ state. The STZ has made a single forward or backward step during its lifetime. The rates for these transitions are contained in the equations of motion, Eq.(\ref{ndot}), for the densities of $\pm$ STZ's, $n_{\pm}(\nu)$. According to these equations, $\rho$ is the rate at which STZ's (of either sign) are annihilated by thermal fluctuations, and $R(\pm s)$ is the rate at which $\mp$ STZ's make transitions to $\pm$ STZ's.  

Now, use these rate factors to compute the  probability of a $\mp$ STZ either being annihilated or making a transition to its $\pm$ state at time $t > 0$, if it formed at $t=0$.  This probability is the normalized waiting-time distribution
\begin{equation}
\Psi_{\mp}(t)= \bigl[\rho + R(\pm s)\bigr] \,e^{-\bigl[\rho + R(\pm s)\bigr]\,t}. 
\end{equation}
The probability that this STZ has {\it not} disappeared or changed its state at time $t$ is
\begin{equation}
\Phi_{\mp}(t) = \int_t^{\infty}dt'\,\Psi_{\mp}(t') = e^{-\,[\rho + R(\pm s)]\,t}.
\end{equation}

Now compute the probability per unit time for a sequence of such transitions.  Suppose, for example, that the STZ  starts in its $-$ state and makes one transition to its $+$ state before being annihilated at time $t_f$.  The probability per unit time for this particular sequence of transitions is 
\begin{equation}
\rho\,\int_0^{\infty} dt_f\,\int_0^{t_f} dt_1\,\Phi_+(t_f - t_1)\,R(+s)\,\Phi_-(t_1)\,\rho\,e^{-\,e_Z/\theta},
\end{equation}
where $\rho\,\exp\,(-\,e_Z/\theta)$ is the formation rate.  For any such sequence of  transitions, the rate is a convolution of functions $\Phi_{\pm}(t_i-t_j)$, multiplied by factors $R(\pm s)$, and integrated over intermediate times $t_1,\,t_2$, etc..  For simplicity, assume that $s$ is not a function of time.  Then we can perform all of these integrations by using the Laplace transforms
\begin{equation}
\tilde \Phi_{\pm}(u) = \int_0^{\infty}\,\Phi_{\pm}(t)\,e^{-\,u\,t}\,dt= {1\over u+\rho+R(\mp s)},
\end{equation} 
so that the Laplace transform of the rate for any sequence of transitions is just a product of factors $\tilde\Phi_{\pm}(u)$ and $R(\pm s)$.  

The sum of all sequences in which the STZ flips forward and backward, any number of times, but always returns to its initial orientation, is 
\begin{equation}
\tilde G(u) = \left[1 - R(+s)\,R(-s)\,\tilde\Phi_+(u)\,\tilde\Phi_-(u)\right]^{-1}. 
\end{equation}
Thus, the Laplace transform for the elementary $-\to +$ or $+ \to -$ process is 
\begin{equation}
\tilde D_{\mp}(u,s) = \rho\,\tilde\Phi_{\pm}(u)\,G(u)\,R(\pm s)\,\tilde\Phi_{\mp}(u)\,\rho\,e^{-\,e_Z/\theta}.
\end{equation}
To compute the jump rate relevant to stress-free, STZ-induced diffusion, set $s=0$ in this formula and integrate over $t_f$ by setting $u=0$.  The result is
\begin{equation}
\tilde D_{\mp}(0,0) = {\rho\,e^{-e_Z/\theta}\,R(0)\over \rho + 2\,R(0)} = {1\over 2}\,{\rho\,e^{-e_Z/\theta}\,\nu\over \rho + \nu},
\end{equation}
where, as usual, $2\,R(0) = 2\,{\cal C}(0) = \nu$.
Denote the total event rate for a given $\nu$ by
\begin{equation}
\label{taudef}
{1\over \tau(\nu)}\equiv 2\,\tilde D_{\mp}(0,0)= {\rho\,e^{-e_Z/\theta}\,\nu\over \rho + \nu},
\end{equation}
which recovers Eq.(\ref{tau}) in the context of diffusion.

Note what is happening here.  If $\nu \gg \rho$, then the STZ makes multiple back and forth transitions during its lifetime, and $\tau(\nu)^{-1} \approx \rho\,\exp\,(-\,e_Z/\theta) = \tau_{\alpha}^{-1}$.  At the opposite extreme, where the internal transitions are very slow so that $\nu \ll \rho$, then $\tau(\nu)^{-1} \approx \nu\,\exp\,(-\,e_Z/\theta) \ll \tau_{\alpha}^{-1}$. In other words, if the diffusing particle encounters a very slow STZ, it most likely will not make a jump at all.

The stress-driven deformation rate, say $D^{pl}(s)$, is the difference between the contributions from sequences with one net forward transition, starting from a $-$ state, and sequences with one net backward transition, starting from a $+$ state:
\begin{eqnarray}
\nonumber
\label{Ddef}
D^{pl}(s)&=& \tilde D_-(0,s) - \tilde D_+(0,s)\cr \\ &=& {2\,\rho\,e^{-e_Z/\theta}\,{\cal C}(s)\,{\cal T}(s)\over \rho + 2\,{\cal C}(s) },
\end{eqnarray}
where
\begin{equation}
{\cal C}(s) = {1\over 2}\,\bigl[R(s) + R(-s)\bigr].
\end{equation}
In the limit of small stress $s$, this result recovers Eq.(\ref{gammadot}) and the viscosity calculation leading to Eq.(\ref{eta}).

\begin{acknowledgments}

I thank Eran Bouchbinder, Michael Cates, Mark Ediger, and Takeshi Egami for helpful discussions during the course of this investigation. This work was supported in part by the Division of Materials Science and Engineering, Office of Basic Energy Sciences, Department of Energy, DE-AC05-00OR-22725, through a subcontract from Oak Ridge National Laboratory. 

\end{acknowledgments}

\end{document}